\documentclass{XrU2005}
\usepackage{epsfig}

\title{Unveiling the distribution of absorption in the AGN population}
\author{T. Dwelly}
\author{M.J. Page}
\affil{Mullard Space Science Laboratory, UCL, Holmbury St. Mary, Dorking, Surrey, UK}

\begin{document}

\maketitle

\begin{abstract}
\vspace{-0.5cm}
We use the very deep {\em XMM-Newton} observations in the CDF-S to measure the
distribution of absorption in the AGN population. We describe
the Monte Carlo method used to unveil the intrinsic properties of the
AGN using their multi-band X-ray colours. The measured
distribution of AGN in $z$, $L_X$ and $N_H$ space is compared with the 
distributions predicted by a number of XLFs and absorption models. 
In contrast to other studies,
we do not find any evidence that the absorption distribution is
dependent on redshift or intrinsic luminosity.
\end{abstract}

\section{Introduction}
\vspace{-0.5cm}
A large population of absorbed AGN are required to explain the hard
spectrum of the extragalactic X-ray background.  However, even though
absorbed AGN outnumber their unabsorbed brethren by a factor of four
in the local Universe (e.g. Risaliti et~al.,1999,ApJ,522,157), the
demographics of the population at higher redshifts are still poorly
understood.  The simple ``unified'' model of AGN attributes this
absorption to a 1--100pc scale dusty torus surrounding the central
black hole and accretion disk, where the degree of absorption seen by
an observer is determined primarily by the angle at which an AGN is
viewed.  By measuring the distribution of absorption in the AGN
population we can place constraints on the typical geometry of such a
torus.  In addition, we can test for the existence of a correlation
between absorption and redshift and/or luminosity.  Therefore, we have
examined a sample of X-ray sources detected in the deep {\em XMM-Newton}
observations of the {\em Chandra} Deep Field-South (CDF-S).  

\section{Observations and sample selection}
\vspace{-0.5cm} 
The {\em XMM-Newton} data in the CDF-S consist of
500ks of observations, of which around 350ks is unaffected by
background flares.  This is the 2nd deepest {\em XMM-Newton} survey to
date. We subdivide the full energy range of the EPIC cameras into four
bands: 0.2--0.5, 0.5--2, 2--5, and 5--10 keV. We detect 299 reliable
X-ray sources to a limiting flux of $\sim 10^{-16}$ erg cm$^{-2}$
s$^{-1}$ in the 0.5--2 keV band. The entire field is covered by deep
{\em Chandra} imaging (Extended-CDF-S project; Lehmer
et~al., 2005, AJ, 129, 1), and the central region is covered by
very deep (1Ms) {\em Chandra} observations (Giacconi et~al., 2002, ApJS, 139, 639).
For unambiguous optical identification, we adopt the high accuracy Chandra
positions where possible.
Extensive spectroscopic identification programs have been carried out
in the CDF-S especially for optical counterparts of the {\em Chandra}
sources, (e.g. Szokoly et~al., 2004, ApJS, 155, 271; Zheng et ~al.,
ApJS, 155, 73Z).  Almost the entire {\em XMM-Newton} field is
also covered by the COMBO-17 survey (Wolf et~al., 2004, A\&A, 421,
913), which provides good photo-z estimates for galaxies to R=24.
We find that 86\% (258/299) of the {\em XMM-Newton} detections have identified optical
counterparts. Of these, 16 sources are stars, 
and 5 suffer from source confusion; these sources are removed from the sample.
\vspace{-0.2cm}
 
\section{Data analysis and results}
\vspace{-0.5cm}
We have used detailed simulations in order to understand the selection
function and any biases inherent to the {\em XMM-Newton} observations (and
the source searching algorithm).  We couple X-ray luminosity functions
with an empirical absorption distribution model to generate a random
synthetic population of ``input'' AGN. The multi-band count-rates expected 
for each simulated AGN are then calculated according to an absorbed power-law plus
reflection spectral model, taking into account the EPIC
response. Simulated multi-band images are generated using the EPIC PSF
together with the exposure maps and unresolved background of the real
observations.  We apply the source searching routine that is used on
the real data in order to find ``output'' sources in these images. Each
``output'' source is matched to the appropriate ``input'' counterpart,
allowing its output parameters (i.e. multi-band count rates), to be related to
input parameters (i.e. $z$, $L_X$, and $N_H$). The simulation process is
repeated for the equivalent of 2000 fields in order to populate densely 
$z$,$L_X$,$N_H$ space. 
See  Loaring et~al. (2005, MNRAS, 362, 137),
and Dwelly et~al. (2005, MNRAS, 360,1426), 
for a detailed description of this process. 
We use this library of simulated sources to find the most likely $N_H$
of the AGN in our CDF-S sample.  For each AGN, we select all those
simulated sources having similar multi-band hardness ratios ($HR$s),
redshifts and full band count rates. The best estimate for the $N_H$
of the real AGN is taken to be the modal ``input'' $N_H$ of this
subset of simulated sources.  We have checked the fidelity of this
method by using it to recover the known $N_H$ of simulated sources. We
find that for simulated sources at $z<2$, and with $N_H>10^{21.5}$
cm$^{-2}$, we are able to recover the input absorption reliably. At
higher redshifts, absorption is shifted out of band, and so we are
only able to place upper limits for moderately absorbed AGN. We use a
similar method to deduce the ``de-absorbed'' intrinsic rest frame
2--10 keV luminosities of the sample. As before, we select all
simulated sources having similar $HR$s, redshift, and full band count
rate to the real AGN. The weighted median luminosity of these
simulated objects is taken to be the best estimate of the intrinsic
luminosity of the AGN. The scatter of the recovered values about the
input values is typically less than 0.2dex. We compare the
$z$,$N_H$,$L_X$ distribution of the CDF-S sample with those predicted
by a number of model $N_H$ functions, coupled with the model XLF of
Ueda et~al. (2003, ApJ, 598, 886).  We have compared the predictions
of a range of absorption distributions including those of Ueda
et~al. (2003), Treister et~al. (2004,ApJ,616,123), and Gilli et~al.
(2001,A\&A,366,407), as well as a simple parameterisation where
$dN/d$log$N_{H}\propto($log$N_H)^{\beta}$, $\beta$~=2,~5,~or~8.
We fold these model populations through our 
Monte Carlo process, and then estimate the output $N_H$ and $L_X$ 
of the simulated sources in the same way as for the AGN in the CDF-S sample. 
The $N_H$ and $L_X$ distributions found
in the CDF-S sample are compared to the predictions made by the best
matching model ($\beta=5$) in figures 1 and 2.

\begin{figure}
\includegraphics[angle=270,width=8cm]{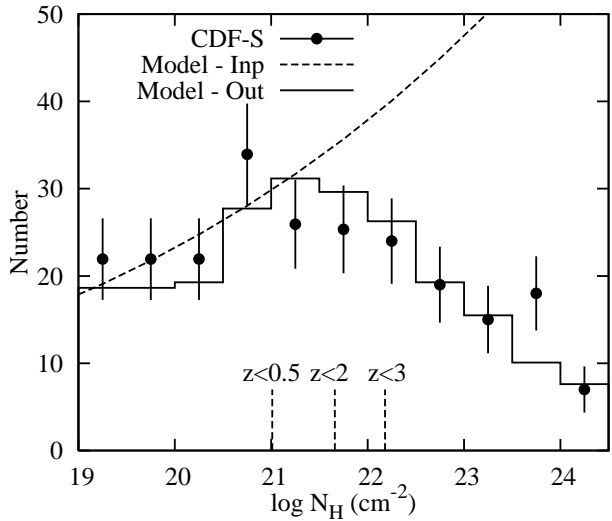}
\caption{The distribution of absorption in the {\em XMM-Newton} CDF-S sample,
compared to the prediction from the best fitting $N_H$
distribution model (the $\beta = 5$ absorption distribution, see text).
We show the input model $N_H$ distribution (before selection effects)
as well as the recovered distribution in the simulated output population.  
Vertical dashed lines show the lowest value at which we can determine the absorbing column at a given redshift. 
All those sources which are determined to have a column 
lower than the limit for their redshift are divided equally between the bins lower than this $N_H$ limit.
}
\label{fig1}
\end{figure}

\begin{figure}
\includegraphics[angle=270,width=8cm,angle=0]{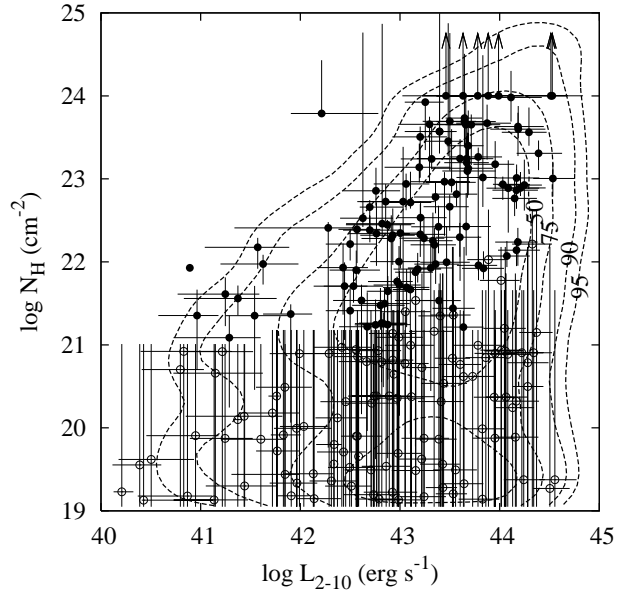}
\caption{
The 2--10 keV $L_X$ vs $N_H$ distribution of the CDF-S sample
(points). Sources for which we can only determine upper limits on
$N_H$ are shown with open symbols. Contours show the distribution predicted
by the $\beta = 5$ model (contour levels are set to
include 50\%,75\%,90\% and 95\% of the model sources).
}
\label{fig2}
\end{figure}

\section{Discussion}
\vspace{-0.5cm}
We have demonstrated a Monte Carlo method for accurately estimating the X-ray
absorption and intrinsic luminosity in a sample of optically identified AGN 
observed with {\em XMM-Newton}. Our CDF-S sample reaches to high redshifts
($z<4$), and spans the knee of the luminosity function ($L^{*} \sim 10^{44}$
erg s$^{-1}$). We find that $\sim 35$\% of the identified sources are
heavily absorbed AGN (log$N_H>22$). Our $N_H$ estimation method finds evidence for 
moderate absorption ($21<$log$N_H<22$) in $\sim70$ AGN; 
Obscuration models where AGN are surrounded by uniformly dense
tori do not predict large numbers of these intermediately absorbed
objects. The model which best reproduces the $N_H$ distribution of the
CDF-S sample is a simple parametrisation, where the number of AGN having
absorption per unit log$N_H$ is proportional to (log$N_H$)$^5$ (the
$\beta=5$ model). After allowing for selection effects in the sample,
we see no strong dependence of the $N_H$ distribution on either
redshift or luminosity in the ranges probed by this sample.  When
compared to the three dimensional $z$,$L_X$,$N_H$ distribution of the
sample, the best matching $N_H$ distribution was again the $\beta=5$
model, which matched with 3-D KS probability P$_{3D-KS}$ = 0.01. The
main source of the disparity is the redshift distributions: the sample
is strongly peaked at $z \sim 0.7$ (e.g. Gilli
et~al. 2003,ApJ,592,721), a feature not reproduced by the model XLF.
This work highlights the importance of high quality, broad-band X-ray
spectral information in determining the $N_H$ distribution of faint
sources, and the need to account rigorously for selection effects.

\end{document}